\documentclass[twocolumn,showpacs,prb]{revtex4}

\usepackage{makeidx}
\usepackage{revsymb}
\usepackage{epsfig}
\usepackage[toc]{appendix}
\usepackage{graphics}
\usepackage{tabularx}
\usepackage{supertabular}

\begin{document}

\title{ SELF-CONSISTENT RPA BASED ON A MANY-BODY VACUUM }
\author{ Mohsen Jema\"{\i}$^{1}$ 
\thanks{email: jemai@ipno.in2p3.fr} 
and 
Peter Schuck$^{2,3}$ 
\thanks{email: schuck@ipno.in2p3.fr} }
\affiliation{$^{1}$ {
\small D\'epartement de Physique, Facult\'e des Sciences de Tunis,
Universit\'e Tunis El Manar, 2092 El Manar, Tunis, Tunisie} \\
$^{2}${\small Institut de Physique Nucl\'eaire d'Orsay, Universit\'e
Paris-Sud, CNRS--IN2P3 15, Rue Georges Clemenceau, 91406 Orsay Cedex, France 
} \\
$^{3}${\small Laboratoire de Physique et Mod\'elisation des Milieux
Condens\'es (LPMMC) (UMR 5493), Maison Jean Perrin, 25 avenue des Martyrs BP 166, 38042 Grenoble cedex 9, France } }
\date{\today }

\begin{abstract}
Self-Consistent RPA is extended in a way so that it is compatable with a variational ansatz for the ground state wave function as a fermionic
many-body vacuum. Employing the usual equation of motion technique, we
arrive at extended RPA equations of the Self Consistent RPA structure. In principle the Pauli principle is, therefore, fully respected. However, the correlation functions entering the RPA matrix can only be obtained from a systematic expansion in powers of some combinations of RPA amplitudes. We demonstrate for a model case that this expansion may converge rapidly.
\end{abstract}

\maketitle


\section{Introduction}

The many body problem still being to a great deal an unsolved problem, new vistas are welcome. We here present an extension of Self-Consistent RPA \cite{art1} which is based on a fermionic many body wave function being the vacuum to an extended RPA destruction operator. Let us expose the main ideas before entering the details of the theory and making an application to a model case. In this work we will consider interacting fermions, however, the extension to the bosonic case is possible. \newline
Many-body theories have a clear cut hirarchy: one body, two body, ..., etc. One body theories are mean field theories. One can write down a variational wave function for the ground state 
\begin{equation}
|Z^{(1)}\rangle =e^{\sum_{ph}z_{ph}^{(1)}c_{p}^{\dagger}c_{h}}|\phi \rangle
\label{eq1}
\end{equation}
\noindent where $z_{ph}^{(1)}$ are variational parameters and $|\phi \rangle =\prod_{h}c_{h}^{\dagger}|vac\rangle $ is an arbitrary Slater determinant built from an antisymmetric product of $N$ fermion creators in the lowest levels, the so-called hole (h) levels below the Fermi energy. One can explicitly evaluate the expectation value of, e.g. a two body Hamiltonian 
\begin{equation}
E_{0}[z^{(1)}]=\frac{\langle Z^{(1)}|H|Z^{(1)}\rangle }{\langle
Z^{(1)}|Z^{(1)}\rangle }  \label{eq2}
\end{equation}
\noindent and minimise with respect to the parameters $z_{ph}^{(1)}$. This leads to the well known Hartree-Fock (HF) equations. That this program can be pulled through hinges to a large extent on the fact that $|Z^{(1)}\rangle $ is the vacuum of quasiparticle operators $a_{p}\propto exp(-Z^{(1)}) c_{p} exp(Z^{(1)}) = c_{p} + \sum_{h} z^{(1)}_{ph} c_{h}$ and similiar for $ a_{h}^{\dagger}$, with $a_{h}^{\dagger}|Z^{(1)}\rangle =a_{p}|Z^{(1)}\rangle =0$. The scheme clearly is of the Raleigh-Ritz (RR) form stating that the so obtained energy is an upper bound to the true ground state energy. \newline
Unfortunately, on the next level, i.e. the two body correlations, there does not exist such a nice closed formalism with RR character. As a consequence several approaches to two body and higher order correlation functions have been elaborated. Some try to be RR variational, some not. For example, the so-called hypernetted chain or Jastrow approach puts a local, usually translational invariant, two body operator in the exponent, with $u(\mathbf{r}-\mathbf{r}^{\prime }) $ the corresponding variational parameters. This leads to a chain of equations, the solution of which may eventually converge to the exact RR variational solution of the corresponding energy, depending on at which level the chain is broken off \cite{art2}. The Gutzwiller ansatz \cite{art3} is taylored to the electron gas on a lattice and puts another two body operator in the exponent penalising double occupancies. However,
the variation usually cannot be pulled through exactly. Most of the time one is satisfied with the approximate solution given by the so-called Gutzwiller equations \cite{art4}. Coupled Cluster Theory (CCT) is an approach \cite{art5} starting also with a two body operator in the exponent (plus eventually higher order operators) but does not lead to a RR scheme, at least not in its original form. Other non RR schemes are based on the BBGKY hirarchy of time dependent density matrices (TDDM). On this subject, there have been some interesting developments, recently \cite{art6}. In the past, several authors, including ourselves, tried to improve RPA as much as possible to overcome certain deficiencies for the calculation of two body correlation functions. It is, e.g., well known that standard RPA (s-RPA) yields two body correlations functions which violate the Pauli principle due
to the so-called quasi-boson approximation inherent to s-RPA (we define
s-RPA as being the small amplitude limit of TDHF implying that the exchange term is included). Usually this leads to a quite strong overbinding of the ground state energy whereas excitation energies may be more correct. We here will formulate the RPA as a diagonalisation problem as is most appropriate for finite Fermi systems, since a good part of the spectrum consists of discrete levels. In this way one usually writes the RPA excitation operator as \cite{art7} 
\begin{equation}
Q_{\nu }^{+}=\sum\limits_{ph}(X_{ph}^{\nu}a_{p}^{\dagger}a_{h}-
Y_{ph}^{\nu} a_{h}^{\dagger} a_{p})  \label{eq3}
\end{equation}
\noindent with $k=(p,h)$ the (particle, hole) levels. The excited state is then written as 
\begin{equation}
|\nu \rangle = Q_{\nu }^{+}|0\rangle  \label{eq4}
\end{equation}
\noindent where $|0\rangle $ is the ground state so that it is the vacuum to the destructors $Q_{\nu }$, i.e. 
\begin{equation}
Q_{\nu }|0\rangle =0~~~~~~ \forall ~\nu \mbox{.} \label{eq5}
\end{equation}
Since this scheme cannot be pulled through in general, in s-RPA the
so-called quasi boson approximation is introduced replacing the fermion pair operators by ideal Bose operators \cite{art7} 
\begin{equation}
a_{p}^{\dagger}a_{h}\rightarrow B_{ph}^{+}  \label{eq6}
\end{equation}
\noindent with 
\begin{equation}
\left[ B_{ph},B_{ph}^{+}\right]=\delta _{pp^{\prime }}\delta _{hh^{\prime }}\mbox{.}
\label{eq7}
\end{equation}
\noindent In this way, (\ref{eq3}) boils down to a Bogoliubov transformation among ideal bosons with the usual HFB vacuum \cite{art7} 
\begin{equation}
|0\rangle \rightarrow e^{(1/4) \sum
z_{p_{1}p_{2}h_{1}h_{2}}^{(B)}B_{p_{1}h_{1}}^{+}B_{p_{2}h_{2}}^{+}}|HF)
\label{eq8}
\end{equation}
\noindent with $a_{h}^{\dagger} a_{p}|HF\rangle \rightarrow B_{ph}|HF)=0$. Since a fermion Hamiltonian can be expanded via Holstein-Primakoff or Belyaev-Zelevinsky boson expansion in powers of the boson operators \cite{art7}, the RPA ends up with a standard HFB approach for interacting bosons. The scheme approaches, of course, a RR variational procedure only in as much as the boson expansion of the Hamiltonian converges. \newline
In the past, we used the RPA operator (\ref{eq3}) with the fermion pair
operators, supposing that the killing condition $Q_{\nu}|0\rangle =0$
existed. However, a corresponding vacuum can only be constructed in some very simplified model cases (see, e.g. \cite{art8}) and, thus, in general, the hypothesis of the existence of a vacuum remained an approximation what perturbed the internal consistency of the so-called Self Consistent RPA (SCRPA) approach \cite{art1, art8, art9}. In this work we will adopt the strategy to keep in (\ref{eq8}) instead of the boson operators the original fermion pair operators. We then will see that this ground state still is vacuum to a generalised RPA operator containing the standard operator (\ref{eq3}) plus, in addition, specific two body operators correcting for the Pauli principle. With that RPA operator we will set up the familiar equation of motion (EOM) method \cite{art7}. The problem will be to evaluate the expectation values of the commutators, since the new RPA operator is non linear, i.e. not of the form of a canonical transformation. Therefore, an
effective systematic expansion scheme will be set up for these expectation values. The procedure will be applied to the very schematic two level Lipkin model where it is shown that the expansion converges fast, even for very small particle numbers.

\section{General Theory}

As mentioned in the introduction, we will consider the following variational ansatz for the many body ground state 
\begin{equation}
|0 \rangle \equiv |Z^{(2)}\rangle = e^{Z^{(2)}}|HF\rangle  \label{eq9}
\end{equation}
\noindent with 
\begin{equation}
Z^{(2)}=\frac{1}{4} \sum z_{p_{1}p_{2}h_{1}h_{2}}^{(2)} a_{p_{1}}^{\dagger}
a_{h_{1}} a_{p_{2}}^{\dagger} a_{h_{2}}  \label{eq10}
\end{equation}
\noindent where $z_{p_{1}p_{2}h_{1}h_{2}}^{(2)}$ are the variational
parameters. In general they can be complex. In principle we can add a one body operator as in (\ref{eq1}) to the exponent and vary $z^{(1)}$ and $z^{(2)}$ simultaneously. In order to keep the formalism as transparent as possible, we will refrain from this complication here and suppose that we work in the standard HF basis. The first task will be to find an RPA operator $Q_{\nu }^{+}$ which applied to $|Z^{(2)}\rangle $ gives the excited states of the system: $|\nu \rangle =Q_{\nu }^{+}|Z^{(2)}\rangle $ and whose hermitian conjugate kills the ground state acting, therefore, as a vacuum 
\begin{equation}
Q_{\nu }|Z^{(2)}\rangle =0~~~~~~ \forall ~\nu \mbox{.}  \label{eq11}
\end{equation}
\noindent We make the following ansatz for the destruction operator\cite{art10} 
\begin{eqnarray}
Q_{\nu }=\sum\limits_{ph} \left[ X_{ph}^{\nu} a_{h}^{\dagger} a_{p} -
Y_{ph}^{\nu} a_{p}^{\dagger}a_{h}\right]  \nonumber \\
+ \frac{1}{2}\sum\limits_{php_{1}p_{2}} \eta_{p_{1}pp_{2}h}^{\nu}
a_{p_{2}}^{\dagger}a_{p_{1}}a_{p}^{\dagger}a_{h}  \nonumber \\
-\frac{1}{2}\sum\limits_{phh_{1}h_{2}} \eta_{ph_{1}hh_{2}}^{\nu}
a_{h_{1}}^{\dagger}a_{h_{2}}a_{p}^{\dagger}a_{h}  \label{eq12}
\end{eqnarray}
\noindent with 
\begin{equation}
Y_{ph}^{\nu}=\sum\limits_{p_{1}h_{1}}X_{p_{1}h_{1}}^{
\nu}z_{p_{1}ph_{1}h}^{(2)}  \label{eq13}
\end{equation}
\noindent and 
\begin{equation}
\eta_{p_{1}pp_{2}h}^{\nu}=\sum\limits_{h_{1}}X_{p_{1}h_{1}}^{
\nu}z_{pp_{2}hh_{1}}^{(2)},  \label{eq14}
\end{equation}
\begin{equation}
\eta_{ph_{1}hh_{2}}^{\nu}=\sum\limits_{p_{1}}X_{p_{1}h_{1}}^{
\nu}z_{pp_{1}hh_{2}}^{(2)}\mbox{.}  \label{eq15}
\end{equation}
\noindent We indeed verify that $Q_{\nu }$ kills the ground state (\ref{eq9}). The amplitudes $X,Y$ can be normalised according to $\langle 0|[Q_{\nu }, Q_{\nu}^{+}]|0\rangle =1$. Evidently (\ref{eq12}) contains the usual RPA operator (\ref{eq3}) plus extra specific 2-body operators. These extra terms serve to fullfil the Pauli principle. They make, however, the calculus rather heavy. In the following, we shortly outline our approach and then give an application to a schematic model. \newline
In order to determine the amplitudes $X,Y$, we proceed as usual and minimisethe following mean excitation energy, equivalent to the minimisation of an (symmetrised) energy weighted sum rule \cite{art9}, 
\begin{equation}
\Omega _{\nu }=\frac{\langle 0|[Q_{\nu },[H,Q_{\nu }^{+}]]|0\rangle +\langle 0|[[Q_{\nu },H],Q_{\nu }^{+}]|0\rangle }{2\langle 0|[Q_{\nu}
,Q_{\nu}^{+}]|0\rangle }\mbox{.}  \label{eq16}
\end{equation}
\noindent Using in this expression the standard RPA operator 
$ Q_{\nu}^{+}=\sum X_{ph}^{\nu }a_{p}^{\dagger}a_{h}-Y_{ph}^{\nu} a_{h}^{\dagger}a_{p}$, we obtain from the minimisation of (\ref{eq16}) with respect to the amplitudes $X$ and $Y$ the usual structure of the (Self Consistent) RPA equations. However, with (\ref{eq8}) the RPA equations become more complicated. Among other things, this stems from the fact that the norm matrix is now non diagonal, i.e. the structure of the equations is as follows 
\begin{equation}
\left( \begin{array}{cc}
R_{11} & R_{12} \\ 
R_{21} & R_{22}
\end{array}
\right) \left(\begin{array}{c}
X \\ Y
\end{array}
\right) =\Omega _{\nu } \left( 
\begin{array}{cc}
N_{11} & N_{12} \\ 
N_{21} & N_{22}
\end{array}
\right) \left( 
\begin{array}{c}
X \\ Y
\end{array}\right)  \label{eq17}
\end{equation}
\noindent with obvious definitions of the matrices $R$ and $N$. They are hermitian matrices by construction, i.e. $R=R^{+}$ and $N=N^{+}$. In order to arrive at the usual structure, we write $N= \tilde{N}^{1/2}N_{0}\tilde{N}^{1/2}$ with 
\begin{equation}
N_{0}=\left( \begin{array}{cc}
1 & 0 \\ 
0 & -1
\end{array}\right)  \label{eq18}
\end{equation}
\noindent and obtain in short hand notation 
\begin{equation}
\tilde{R}\tilde{\chi}=\Omega _{\nu }N_{0}\tilde{\chi}  \label{eq19}
\end{equation}
\noindent with $\tilde{R}=\tilde{N}^{-1/2}R\tilde{N}^{-1/2}$ and 
$\tilde{\chi} =\tilde{N}^{1/2}\chi $. This equation now has globally the same mathematical structure as standard RPA or SCRPA (details are different, of course). In particular, we have for the normalisation 
$\tilde{X}^{2}-\tilde{Y }^{2}=1$. Of course, the task is not finished. The difficulty resides in how to express the correlation functions in $\tilde{R}$ and $\tilde{N}$ by the amplitudes $\tilde{X},\tilde{Y}$, since the form (\ref{eq12}) of the RPA operator, as already mentioned, does not constitute a canonical transformation among operators and can, therefore, not be inverted in a straightforward manner. We will set up an expansion scheme which can be paraphrased as follows: Suppose we did not know how to invert the Bogoliubov transformation from ordinary to quasiparticles in standard BCS theory. How to calculate expectation values in the BCS state nontheless? Among other ways, this can go as follows. As is well known, the BCS state can be written as a coherent state: 
$|BCS\rangle =exp[\sum z_{k}c_{k}^{\dagger}c_{-k}^{\dagger}]|vac\rangle $. We want to calculate, e.g. the pairing tensor $\kappa_{k}=\langle BCS|c_{-k}c_{k}|BCS\rangle/\langle BCS|BCS\rangle $. We have $c_{k}|BCS\rangle =z_{k}c_{-k}^{\dagger}|BCS\rangle $. Multiplying this relation with $c_{-k}$ and repeating the operation leads to the well known BCS expression $\kappa_{k}=u_{k}v_{k}$ where we supposed for simplicity $z_{k}$ to be real, i.e. 
$\langle c_{-k}c_k\rangle =\langle c^\dagger_k c^\dagger_{-k}\rangle $, and used $z_{k}=v_{k}/u_{k}$ with $u_{k}^{2}+v_{k}^{2}=1$. Using an analogous method to calculate the correlation functions in the RPA matrix leads to a series which, unfortunately, does not break off after a low number of terms unless one deals with very few particle states. As the example with the BCS state shows however, the series contains nontrivial terms and may, thus, converge rapidly. If convergence is achieved, Pauli principle is respected fully, since the theory is based on a wave function. Of course, the final success will depend on that supposed rate of convergence. This ends the rough outline of the theory. Let us now make an application to the Lipkin model which often has served in the past as a testing ground for many body theories. We will see how in that model the series expansion works and that in this case the convergence of the series is, indeed, fast and performant.

\section{ Aplication to a schematic model}

In this section we want to apply our theory to the schematic Lipkin model \cite{art11} which is an exactly solvable two level model with separable monopole-monopole interaction. The Hamiltonian is given by \cite{art7} (without loss of generality we put the intershell spacing $\varepsilon = 1$) 
\begin{equation}
H = \hat{J}_{0} - \frac{V}{2}\left( \hat{J}_{+}^{2} + \hat{J}_{-}^{2}\right)
\label{eq20}
\end{equation}
\noindent with $V$ the coupling constant and 
\begin{eqnarray}
\hat{J}_{0} & = & \frac{1}{2}\sum_{m=1}^{\Omega }\left(
c_{1m}^{\dagger}c_{1m}-c_{0m}^{\dagger}c_{0m}\right) \,,  \nonumber \\
\hat{J}_{+} & = & \sum_{m=1}^{\Omega }\;c_{1m}^{\dagger}c_{0m}\,,~~~~~~~~~ 
\hat{J}_{-}=(\hat{J}_{+})^{\dagger }  \label{eq21}
\end{eqnarray}
where $\Omega $ is the degeneracy of one shell. We take $\Omega = N $ with $N$ the particle number, i.e. the lower shell is filled for $V = 0$. The SU2 commutation relations are 
\begin{equation}
\left[ \hat{J}_{+},\hat{J}_{-}\right] =2\,\hat{J}_{0}\;,~~~~~~~~~ 
\left[\hat{J}_{0},\hat{J}_{\pm }\right] =\pm \,\hat{J}_{\pm }\mbox{.}\label{eq22}
\end{equation}
The variational ground state corresponding to (\ref{eq9}) is given by 
\begin{equation}
|Z\rangle =e^{\hat{Z}}|HF\rangle  \label{eq23}
\end{equation}
\noindent with 
\begin{equation}
\hat{Z}=z\hat{J}_{+}\hat{J}_{+}  \mbox{.}\label{eq24}
\end{equation}
We verify that this correlated state is vacuum to the following destruction operator \cite{art10} 
\begin{equation}
Q=X\hat{J}_{-}-Y\bar{J}_{+}  \label{eq25}
\end{equation}
\noindent with $z=Y/\left( NX\right) $ and 
\begin{equation}
\bar{J}_{+}=\left(1-\eta \hat{J}_{0}\right)\hat{J}_{+}, ~~~~~~ \eta =\frac{2}{N} \mbox{.}  \label{eq26}
\end{equation}
We note that the non-standard correction term is of order $1/N$. In this model the matrix elements of eq.(\ref{eq17}) are given by 
\begin{eqnarray}
R_{11} & = & \frac{1}{2}\left( \langle \left[ \hat{J}_{-},\left[ H , \hat{J}_{+}\right] \right] \rangle +\langle \left[ \left[ \hat{J}_{-},H\right] , \hat{J}_{+}\right] \rangle \right)  \nonumber \\
R_{12} & = & -\frac{1}{2}\left( \langle \left[ \hat{J}_{-},\left[ H, 
\overline{J}_{-}\right] \right] \rangle +\langle \left[ \left[ \hat{J}_{-},H \right] , \overline{J}_{-}\right] \rangle \right)  \nonumber \\
R_{21} & = & -\frac{1}{2}\left( \langle \left[ \overline{J}_{+},\left[ H, \hat{J}_{+}\right] \right] \rangle +\langle \left[ \left[ \overline{J}_{+},H \right],\hat{J}_{+}\right] \rangle \right)  \nonumber \\
R_{22} & = & \frac{1}{2}\left( \langle \left[ \overline{J}_{+},\left[ H, \overline{J}_{-}\right] \right] \rangle +\langle \left[ \left[ \overline{J}_{+},H\right] , \overline{J}_{-}\right] \rangle \right)  \label{eq27}
\end{eqnarray}
and the elements of the norm matrix, 
\begin{eqnarray}
N_{11} & = & \langle \left[ \hat{J}_{-},\hat{J}_{+}\right] \rangle =
-2\langle \hat{J}_{0}\rangle  
\nonumber \\
N_{12} & = & -\langle \left[ \hat{J}_{-},\overline{J}_{-}\right] \rangle =-\eta \langle \hat{J}_{-}^{2}\rangle  
\nonumber \\
N_{21} & = & -\langle \left[ \overline{J}_{+},\hat{J}_{+}\right] \rangle =-\eta\langle \hat{J}_{+}^{2}\rangle  
\nonumber \\
N_{22} & = & \langle\left[\overline{J}_{+},\overline{J}_{-}\right]\rangle 
\nonumber \\
& = & 2 \langle \hat{J}_{0}\rangle - 2 \eta \left[ 2\langle \hat{J}
_{0}\rangle + 2\langle \hat{J}_{0}^{2}\rangle -\langle \hat{J}_{+}\hat{J}_{-}\rangle \right]  
\nonumber \\
& & +\eta^{2}\biggl\{ 2\langle \hat{J}_{0}\rangle + 4\langle \hat{J}
_{0}^{2}\rangle + 2\langle \hat{J}_{0}^{3}\rangle -3\langle \hat{J}_{+}\hat{J}_{-}\rangle  
\nonumber \\
&& ~~~~~~~ -2\langle \hat{J}_{+}\hat{J}_{0}\hat{J}_{-}\rangle \biggr\}\mbox{.}
\label{eq28}
\end{eqnarray}
To calculate the double commutators in (\ref{eq27}), we use the Jakobi
identity $\left[ A,\left[ H, B\right] \right] - \left[\left[ A, H\right], B \right] = \left[ H, \left[ A,B\right] \right] $, and obtain 
\begin{widetext}
\begin{eqnarray}
R_{11} & = & -2\langle \hat{J}_{0}\rangle +V\left( \langle \hat{J}
_{-}^{2}\rangle +\langle \hat{J}_{+}^{2}\rangle \right)  
\nonumber \\
R_{12} & = & R_{21} = -V\biggl\{ 2\langle \hat{J}_{0}\rangle + 4\langle \hat{J}_{0}^{2}\rangle -2\langle \hat{J}_{+}\hat{J}_{-}\rangle 
-\eta \left[ 3\langle \hat{J}_{0}\rangle + 8 \langle \hat{J}_{0}^{2}\rangle + 4\langle \hat{ J }_{0}^{3}\rangle -7\langle \hat{J}_{+}\hat{J}_{-}\rangle -6\langle \hat{J}_{+}\hat{J}_{0}\hat{J}_{-}\rangle 
\right] \biggr\} 
\nonumber \\
R_{22} & = & -2\langle \hat{J}_{0}\rangle 
+V\left( \langle \hat{J}_{-}^{2}\rangle +\langle \hat{J}_{+}^{2}\rangle \right) +\eta \left[ 4\langle \hat{J}_{0}\rangle +4\langle \hat{J}_{0}^{2}\rangle + 2\langle \hat{J}_{+}\hat{J}_{-}\rangle 
+ V \left( \langle \hat{J}_{-}^{2}\rangle +\langle \hat{J}_{+}^{2}\rangle \right)\right]   
\nonumber \\
& & -\eta ^{2}\biggl\{
 2\langle \hat{J}_{0}\rangle +4\langle \hat{J}
_{0}^{2}\rangle +2\langle \hat{J}_{0}^{3}\rangle  
-3\langle \hat{J}_{+}\hat{J}_{-}\rangle 
-2\langle \hat{J}_{+}\hat{J}_{0}\hat{J}_{-}\rangle  
\nonumber \\
& & + V \left[ 
\frac{13}{2}\left( \langle \hat{J}_{-}^{2}\rangle +\langle \hat{J}_{+}^{2}\rangle \right)  
+8\left( \langle \hat{J}_{0}\hat{J} _{-}^{2}\rangle 
+ \langle \hat{J}_{+}^{2}\hat{J}_{0}\rangle \right) 
+\left(\langle \hat{J}_{0}^{2}\hat{J}_{-}^{2}\rangle +\langle \hat{J}_{+}^{2}\hat{J}
_{0}^{2}\rangle \right)
-2\left( \langle \hat{J}_{+}^{3}\hat{J}_{-}\rangle 
+ \langle \hat{J}_{+}\hat{J}_{-}^{3}\rangle \right) \right]
\biggr\} \mbox{.}
\label{eq29}
\end{eqnarray}
\end{widetext}
Diagonalising the norm matrix we arrive at an equation which has the
sturcture (\ref{eq19}), i.e. the same structure as the standard RPA or SCRPA equations. \newline
At this point we shortly would like to explain how one can calculate the correlators appearing in the RPA matrix within our formalism following in spirit the example given in section II for the calculation of the pairing tensor. Let us demonstrate this on the example of $\langle Z| \hat{J}_{0}|Z\rangle/\langle Z|Z\rangle $ to lowest non trivial order. We have 
\begin{eqnarray}
\hat{J}_{0}|Z\rangle & = & \left[ -\frac{N}{2}+2z\hat{J}_{+}^{2}\right]
|Z\rangle  
\nonumber \\
\hat{J}_{-}|Z\rangle & = & 2z(N-1)\hat{J}_{+} |Z\rangle + \ldots  \nonumber \\
\hat{J}_{-}\hat{J}_{-}|Z\rangle & = & 2z(N-1) \left[ -2\hat{J}_{0} + \hat{J}_{+}\hat{J}_{-}\right] |Z\rangle + \ldots  
\nonumber \\
& = & 2 z (N-1)\left[-2\hat{J}_{0}+2z( N-1)\hat{J}_{+} \hat{J}_{+}\right]
|Z\rangle + \ldots  
\nonumber
\end{eqnarray}
Now, we calculate the average value in the state $|Z\rangle $ and use $
\langle Z|\hat{J}_{-}\hat{J}_{-}|Z\rangle =\langle Z|\hat{J}_{+}\hat{J}_{+}|Z\rangle $ 
\begin{eqnarray}
\langle Z| \hat{J}_{0}|Z\rangle & = & -\frac{N}{2}\langle Z|Z\rangle
+2z\,\langle Z|\hat{J}_{+}\hat{J}_{+}|Z\rangle  
\nonumber \\
\langle Z| \hat{J}_{+}\hat{J}_{+}|Z\rangle & = & \frac{-4z(N-1)} {
1-4z^{2}(N-1)^{2}}\langle Z| \hat{J}_{0}|Z\rangle  \mbox{.}
\nonumber
\end{eqnarray}
Thus, 
\begin{equation}
\frac{\langle Z| \hat{J}_{0}|Z\rangle }{\langle Z|Z\rangle } = -\frac{N}{2} \frac{1-4z^{2}(N-1)^{2}}{1-4z^{2}(N-1)(N-3) }\mbox{.} 
\label{eq30}
\end{equation}
With this it becomes evident how to push the expansion to higher order and how to do analogous expansions for all other correlators. A more complete set of correlation functions is given in the Appendix (\ref{app1}). We now will solve eq (\ref{eq17}) for different particle numbers. For $N=2$, eq (\ref{eq17}) reads 
\begin{eqnarray}
& & \left( \begin{array}{cc}
2\left( 1-4z^{2}+4Vz\right) & -V\left( 1+4z^{2}\right) \\ 
-V\left( 1+4z^{2}\right) & 2\left( 1+2Vz\right)
\end{array}\right) \left( 
\begin{array}{c}
X \\ Y
\end{array}\right)  
\nonumber \\
& & = \Omega _{\nu }\left( \begin{array}{cc}
2\left( 1-4z^{2}\right) & 4z \\ 
4z & -2
\end{array}\right) \left( 
\begin{array}{c}
X \\ Y
\end{array}\right)  \label{eq31}
\end{eqnarray}
where we divided by the common normalisation factor $(1+4z^{2})$. Let us begin with the diagonalisation of the norm matrix 
\begin{equation}
n=\left( \begin{array}{cc}
2\left( 1-4z^{2}\right) & 4z \\ 
4z & -2
\end{array}\right)  \mbox{.}\label{eq32}
\end{equation}
Its eigenvalues are 
\begin{eqnarray}
e_{+} & = & 2\sqrt{1+4z^{4}}-4z^{2}>0,  \nonumber \\
e_{-} & = & -2\sqrt{1+4z^{4}}-4z^{2}<0  \label{eq33}
\end{eqnarray}
and the corresponding unitary matrix is given by 
\begin{equation}
U=\left( \begin{array}{cc}
u_{+} & u_{-} \\ 
v_{+} & v_{-}
\end{array}
\right) =\left( \begin{array}{cc}
u & -v \\ 
v & u
\end{array}
\right),~~ U^{+}=\left( \begin{array}{cc}
u & v \\ 
-v & u
\end{array}\right)  \label{eq34}
\end{equation}
with $(\det \left( U\right) =1)$ 
\begin{eqnarray}
u & = & \frac{\left(1-2z^{2}\right) +\sqrt{1+4z^{4}}}
{\sqrt{4z^{2} +\left[1-2z^{2}+\sqrt{1+4z^{4}}\right] ^{2}}},  
\nonumber \\
v & = & \frac{2z}{\sqrt{4z^{2}+\left[1-2z^{2}-\sqrt{1+4z^{4}}\right]^{2}} } \mbox{.}
\label{eq35}
\end{eqnarray}
The norm matrix (\ref{eq32}) can be written as $n=\tilde{n}^{1/2}N_{0}\tilde{n}^{1/2}$ with 
\begin{eqnarray}
\tilde{n}^{1/2}=U^{+}\left( 
\begin{array}{cc}
\sqrt{e_{+}} & 0 \\ 
0 & \sqrt{| e_{-}| }
\end{array}\right) \mbox{.}  \nonumber
\end{eqnarray}
We thus obtain $\tilde{r}\tilde{\chi} = \Omega _{\nu} N_{0} \tilde{\chi}$ or explicitly 
\begin{equation}
\left(\begin{array}{cc}
\tilde{r}_{11} & \tilde{r}_{12} \\ 
\tilde{r}_{21} & \tilde{r}_{22}
\end{array}\right) \left( 
\begin{array}{c}
\tilde{X} \\ \tilde{Y}
\end{array}\right) =\Omega_{\nu }\left( 
\begin{array}{cc}
1 & 0 \\ 
0 & -1
\end{array}\right) \left( \begin{array}{c}
\tilde{X} \\ \tilde{Y}
\end{array}\right)  \label{eq36}
\end{equation}
with $\tilde X^2 - \tilde Y^2 = 1$ and 
\begin{eqnarray}
\tilde{r}_{11} & = & \frac{| e_{-}| }{4}\left(
u^{2}r_{11}+v^{2}r_{22}-2uvr_{12}\right) ,  
\nonumber \\
\tilde{r}_{12} & = & \tilde{r}_{21}=\frac{1}{2}\left( u^{2}-v^{2}\right) r_{12}+ \frac{1}{2}uv\left( r_{11}-r_{22}\right)  
\nonumber \\
\tilde{r}_{22} & = & \frac{e_{+}}{4}\left( u^{2} r_{22} + v^{2} r_{11} +2 uv r_{12}\right)  \mbox{.}\nonumber
\end{eqnarray}
Multiplying first equation of (\ref{eq36}) with $\tilde{Y}$ and second with $\tilde{X}$ we obtain after adding and substracting both equations 
\begin{eqnarray}
& & \left( \tilde{r}_{11}+\tilde{r}_{22}\right) \tilde{X} \tilde{Y} 
+ \tilde{r} _{12}\left( \tilde{X}^{2}+\tilde{Y}^{2}\right) = 0  \nonumber \\
& & \left( \tilde{r}_{11}-\tilde{r}_{22}\right) \tilde{X} \tilde{Y} 
+ \tilde{r} _{12} \left( \tilde{X}^{2}- \tilde{Y}^{2} \right) = 2 \Omega _{\nu }\tilde{X} \tilde{Y}  \mbox{.}\label{eq37a}
\end{eqnarray}
For the solution of this equation, we need to express $\tilde X$ and $\tilde Y $ by $X $, $Y$. This is given below in (\ref{eq38a}). With $z=Y/(NX)$ one then obtains 
\begin{eqnarray}
z_{\pm } & = & \frac{V}{2}\frac{1}{ 1 \pm \sqrt{1+V^{2}} }  
\nonumber \\
\Omega _{\pm } & = & \pm \sqrt{1+V^{2}} \mbox{.}\label{eq37b}
\end{eqnarray}
It can readily be checked that this solution coincides with the exact one. Please notice that the equality $\tilde{r}_{11}=\tilde{r}_{22}$ only holds at the solution (\ref{eq37b}). Therefore, only after convergence to the solution we obtain the usual RPA structure with eigenvalues which come in pairs $\pm |\Omega _{\nu }|$. This also holds for the other cases with $N>2$ considered below. We surmise that this is a general property of our theory. 
\newline
The new amplitudes $\tilde{X} $, $\tilde{Y}$ in terms of $X$, $Y$ are given by 
\begin{equation}
\left( \begin{array}{c}
\tilde{X} \\ \tilde{Y}
\end{array}\right) = \left( 
\begin{array}{cc}
u\sqrt{e_{+}} & -v\sqrt{e_{+}} \\ 
v\sqrt{|e_{-}|} & u\sqrt{|e_{-}|}
\end{array}\right) \left( 
\begin{array}{c}
X \\ Y
\end{array}\right)  \label{eq38a}
\end{equation}
and vice versa 
\begin{equation}
\left( 
\begin{array}{c}
X \\ Y
\end{array}\right) = \left( 
\begin{array}{cc}
\frac{u}{\sqrt{e_{+}}} & \frac{v}{\sqrt{|e_{-}|}} \\ 
-\frac{v}{\sqrt{e_{+}}} & \frac{u}{\sqrt{|e_{-}|}}
\end{array}
\right) \left( 
\begin{array}{c}
\tilde{X} \\ \tilde{Y}
\end{array}\right) \mbox{.} \label{eq38b}
\end{equation}
Therefore, the RPA operator can be written as 
\begin{eqnarray}
Q_{\nu } & = & X\hat{J}_{-} - Y\bar{J}_{+}  
\nonumber \\
& = & \tilde{X} \hat{J}_{1} + \tilde{Y} \hat{J}_{2}
\end{eqnarray}
with 
\begin{eqnarray}
\hat{J}_{1} & = & \frac{1}{\sqrt{e_{1}}} \left( u\hat{J}_{-} - v\bar{J}
_{+}\right)  
\nonumber \\
\hat{J}_{2} & = & \frac{1}{\sqrt{|e_{2}|}}\left( v\hat{J}_{-} + u\bar{J}_{+}\right)  \mbox{.}\nonumber
\end{eqnarray}
The solution for $N=4$, $N=6$ and $N=10$ can be obtained with the
correlation functions given in the appendix (\ref{app1}). The results are shown in Figs \ref{E_1N4}, \ref{E_1N6}, \ref{E_1N10},  and \ref{E_1N20}. We see that SCRPA with $\eta \neq 0$ (SCRPA($\eta $)) systematically and substantially improves over the $\eta = 0$ case (SCRPA(0)) in the range $\chi<1$. This fact shows that the extra correlations introduced in the excitation operator (\ref{eq12}) improving the Pauli principle play an important role . In the range $\chi>1$, ground and excited states become strongly mixed and SCRPA in the 'spherical' basis deviates more and more from the exact solution. The good agreement of SCRPA$(\eta)$ with the exact solution also demonstrates that the convergence of our expansion scheme is fast. In future work it will be very interesting to apply our theory also to the symmetry broken regime. 

\section{Conclusions}

In this work we have set up a Self-Consistent RPA (SCRPA) theory which is based on a fermionic many body vacuum. That is, we constructed a killing operator of the correlated vacuum $|0\rangle \rightarrow
exp[Z^{(2)}]|HF\rangle$, with $Z^{(2)}$ a two body operator, see eq (\ref{eq9}), (\ref{eq10}) which also is the starting point of Coupled Cluster theory \cite{art5}. This killing operator consists of the usual $ph$ RPA operator with, additionally, specific two body terms correcting for the Pauli principle. With this extended RPA operator, we develope the standard Equation of Motion technique and arrive at SCRPA equations of the usual mathematical structure. Since the extended RPA operator is not of the type of a canonical transformation between operators, the evaluation of the correlation functions contained in the RPA matrix in terms of the RPA amplitudes is not evident. We proposed a systematic expansion scheme in terms of higher and higher correlations functions (there may be room for improving this scheme) and applied it to the two level Lipkin model as a testing ground. It is found that the additional terms in the RPA operator systematically and substantially improve the results. We want to stress that once convergence is achieved with the mentioned expansion, the Pauli principle is fully respected, since our method is based on a fermionic many body ground state which is the vacuum to an (extended) RPA operator. In this
way SCRPA which in the past suffered from the fact that it was based on a non existing vacuum is now put on grounds with more internal consistency. However, the practical success will depend on the convergence rate of our expansion scheme in more realistic situations. This is similar to the convergence rate of the hirarchy of hypernetted chain equations with Jastrow theory. In this work we only investigated the symmetry unbroken phase of the Lipkin model. The symmetry broken phase as well as other models for strongly interacting fermions shall be investigated in the future. On the formal level some questions still need to be considered also. The most important one is whether the Goldstone theorem when working in a symmetry broken phase still holds as in s-RPA. In particular this concerns the fullfillment of conservation laws, the appearence of a zero mode and the respect of Ward identities.

\begin{acknowledgments}
We thank J. Dukelsky for very useful discussions. 
A critical reading of the manuscript and interesting comments from M. Holzmann are appreciated.
\end{acknowledgments}

\begin{appendix}

\section{Correlation functions}
\label{app1} 
In this Appendix, we list correlation functions to lowest non-trivial order as they are needed for the RPA calculation 
\[
\frac{\langle Z|\hat{J}_{0}|Z\rangle}{\langle Z|Z\rangle} 
= -\frac{N}{2}\frac{1 + 2(N-1)(N-4) z^{2}}
{1 + 2N(N-1) z^{2}}
\]
\[
\frac{\langle Z|\hat{J}_{0}^{2}|Z\rangle }{\langle Z|Z\rangle}=\frac{N}{4}
\frac{N + 2(N-1)(N-4)^{2} z^{2}}
{1 + 2N(N-1) z^{2}}
\]
\[
\frac{\langle Z|\hat{J}_{0}^{3}|Z\rangle }{\langle Z|Z\rangle}=
-\frac{N}{8}\frac{N^{2} + 2(N-1)(N-4)^{3} z^{2}}
{1 + 2N(N-1) z^{2}}
\]
\[
\frac{\langle Z|\hat{J}_{-}^{2}|Z\rangle }{\langle Z|Z\rangle}
=\frac{\langle Z|\hat{J}_{+}^{2}|Z\rangle }{\langle Z|Z\rangle}=
\frac{ 2N(N-1) z }{1 + 2N(N-1) z^{2}}
\]

\[
\frac{\langle Z|\hat{J}_{+}\hat{J}_{-}|Z\rangle }{\langle Z|Z\rangle}
=\frac{ 4 N (N-1)^2 z^{2}}{1 + 2N(N-1) z^{2}}
\]
\[
\frac{\langle Z|\hat{J}_{+}\hat{J}_{0}\hat{J}_{-}|Z\rangle }{\langle Z|Z\rangle}
=- \frac{4N(N-1)^2(N-2) z^{2}}
{1 + 2N(N-1) z^{2}}
\]
\[
\frac{\langle Z|\hat{J}_{+}^{2}\hat{J}_{0}|Z\rangle }{\langle Z|Z\rangle}
=\frac{\langle Z|\hat{J}_{0}\hat{J}_{-}^{2}|Z\rangle }{\langle Z|Z\rangle}
= -\frac{ N^2( N-1) z }{1 + 2N(N-1) z^{2} }
\]
\[
\frac{\langle Z|\hat{J}_{+}^{2}\hat{J}_{0}^{2}|Z\rangle }{\langle Z|Z\rangle}
=\frac{\langle Z|\hat{J}_{0}^{2}\hat{J}_{-}^{2}|Z\rangle}{\langle Z|Z\rangle}
=\frac{1}{2}\frac{ N^{3}(N-1)z  }
{1 + 2N(N-1) z^{2}}
\]
\[
\frac{\langle Z|\hat{J}_{+}^{3}\hat{J}_{-}|Z\rangle }{\langle Z|Z\rangle}
=\frac{\langle Z|\hat{J}_{+}\hat {J}_{-}^{3}|Z\rangle}{\langle Z|Z\rangle}
= 0
\]
\end{appendix}

 
\begin{widetext}
\begin{figure}[htb]
 \begin{center}
  \epsfig{file=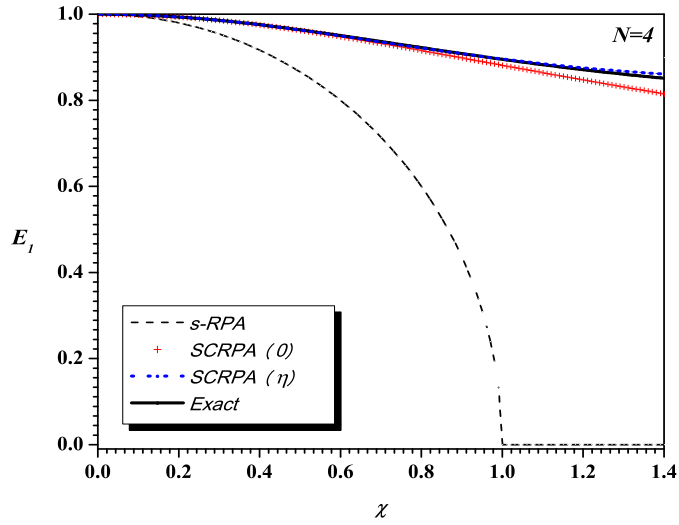,width=10cm,height=8cm}
   \caption{First excited state energy with s-RPA, SCRPA(0), SCRPA($\eta $), and exact solution 
		for $N=4$ as a function of $\chi=V(N-1)$. }
     \label{E_1N4}
      \end{center}
       \end{figure}

\begin{figure}[htb]
 \begin{center}
  \epsfig{file=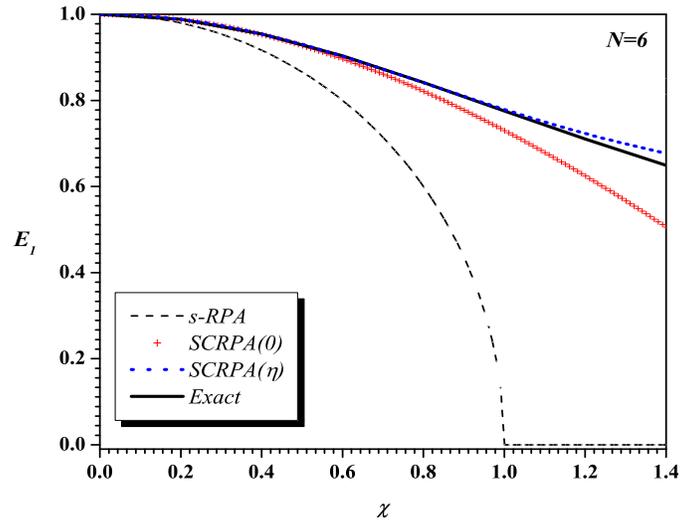,width=10cm,height=8cm}
   \caption{Same as Fig. \ref{E_1N4} but for $N = 6$.}
    \label{E_1N6}
     \end{center}
      \end{figure}

\begin{figure}[htb]
 \begin{center}
  \epsfig{file=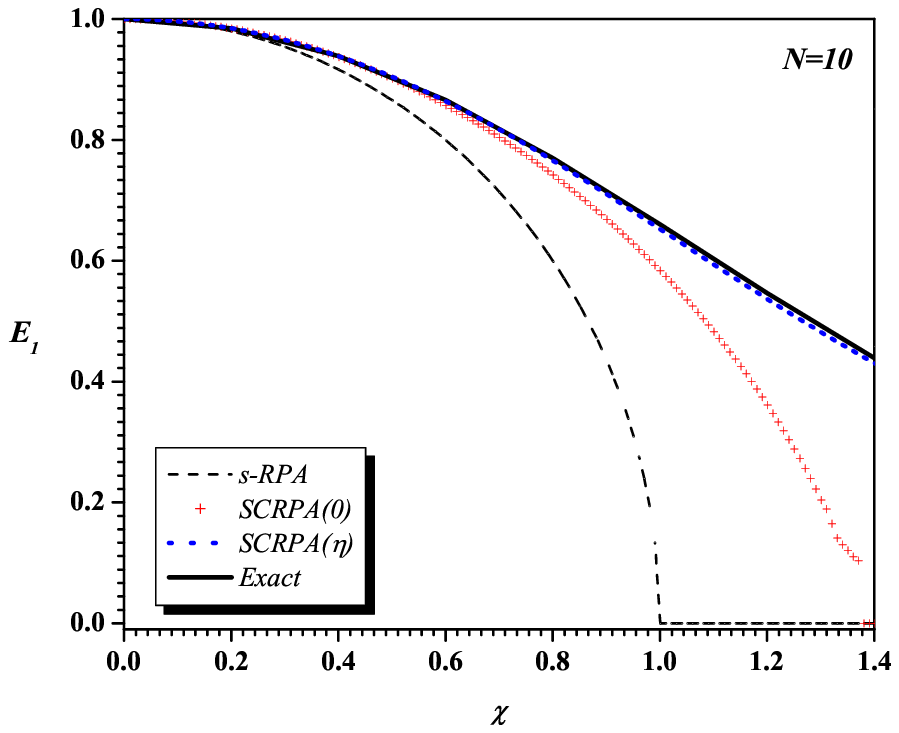,width=10cm,height=8cm}
   \caption{Same as Fig. \ref{E_1N4} but for $ N = 10 $.}
    \label{E_1N10}
     \end{center}
      \end{figure}

\begin{figure}[htb]
 \begin{center}
  \epsfig{file=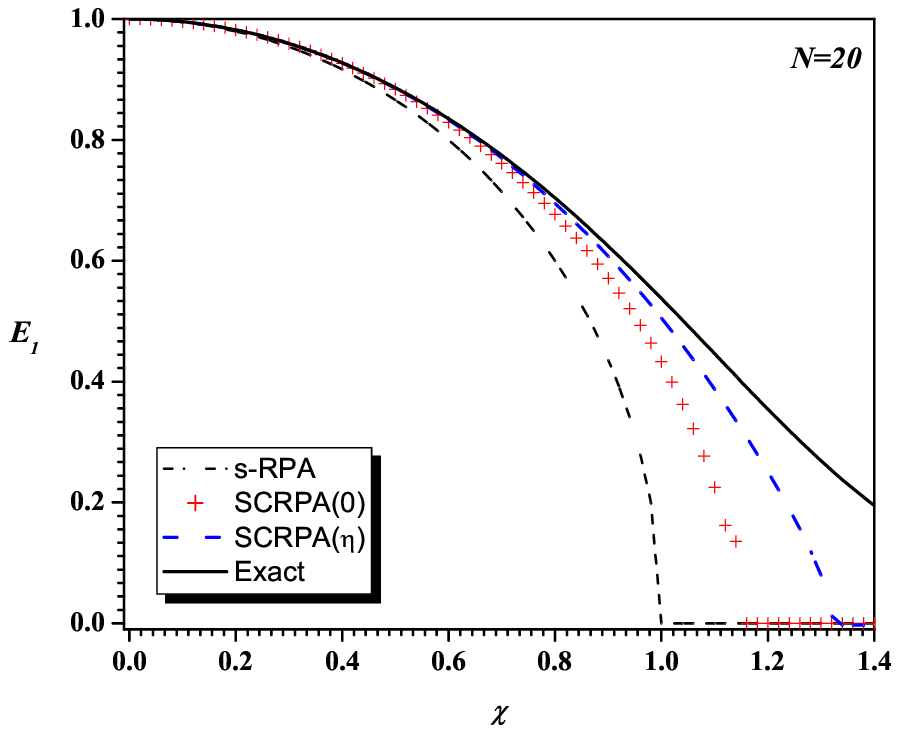,width=10cm,height=8cm}
   \caption{Same as Fig. \ref{E_1N4} but for $ N = 20 $.}
    \label{E_1N20}
     \end{center}
      \end{figure}

 \end{widetext}

\end{document}